\documentclass[aps, prl, reprint, superscriptaddress, nofootinbib, longbibliography, floatfix]{revtex4-2}
\usepackage{graphicx}
\usepackage{bm}
\usepackage{amsmath}
\usepackage{amssymb}
\usepackage{hyperref}
\usepackage{slashed}
\usepackage{url}
\usepackage{orcidlink}

\newcommand{\ST}{\mathcal S_T}
\newcommand{\Bcal}{\mathcal B}
\newcommand{\Bvac}{\mathcal B_0}
\newcommand{\dd}{\mathrm d}

\hypersetup{
colorlinks=true,
linkcolor=blue,
citecolor=blue,
urlcolor=blue
}

\begin{document}


\title{Pseudo-Nambu-Goldstone Boson Motion from Thermal Thresholds}

\author{Mussawir Khan\,\orcidlink{0009-0006-3673-4243}}
\email{mussawirkhan@ihep.ac.cn}
\affiliation{
State Key Laboratory of Particle Astrophysics,
Institute of High Energy Physics,
Chinese Academy of Sciences,
Beijing 100049, China
}
\affiliation{University of Chinese Academy of Sciences, Beijing 100049, China}

\author{Ali Muhammad}
\email{alimuhammad@phys.qau.edu.pk}
\affiliation{%
  CAS Key Laboratory of Theoretical Physics,
  Institute of Theoretical Physics,
  Chinese Academy of Sciences,
  Beijing 100190, China
}
\affiliation{%
  School of Physical Sciences,
  University of Chinese Academy of Sciences,
  Beijing 100049, China
}

\author{Tianjun Li}
\email{tli@mail.itp.ac.cn}
\affiliation{School of Physics, Henan Normal University, Xinxiang 453007, China}

\begin{abstract}
Coherent motion of light scalar fields in the early Universe is typically attributed to an initial displacement, a primordial velocity, or a prescribed temperature-dependent potential. We identify a distinct microscopic mechanism in which differential thermal evolution of multiple non-aligned symmetry-breaking sources rotates the effective vacuum direction of a pseudo-Nambu--Goldstone boson (pNGB), driving motion from an initially stationary minimum. The mechanism is realized in the Majoron sector of a Type-I seesaw mechanism, where heavy neutrinos with separated thermal thresholds supply the required differential thermal evolution. For a representative benchmark, the thermal source direction rotates by $|\Delta\theta_T|\simeq0.76$~rad and drives the Majoron to a peak velocity $|\dot\theta|/H\simeq0.62$, while the field tracks the moving minimum to sub-milliradian accuracy. This establishes a microscopic finite-temperature origin for pNGB motion, with implications for spontaneous leptogenesis and, more generally, for any system with multiple non-aligned symmetry-breaking sources that evolve differently with temperature.
\end{abstract}

\maketitle
\paragraph{Introduction.}
Light scalar fields associated with approximate symmetries are ubiquitous in early-Universe physics. Pseudo-Nambu--Goldstone bosons (pNGBs) arise in axion models \cite{Preskill1983}, relaxion constructions \cite{Graham2015}, dark-energy scenarios \cite{Carroll1998}, and extensions of the Standard Model (SM) with spontaneously broken global symmetries \cite{Chikashige1980,Chikashige1981,GelminiRoncadelli}. Their coherent motion can contribute to dark matter \cite{Turner1986,KimCarosi2010,Marsh2016}, source cosmological charge asymmetries \cite{CohenKaplan1987,CohenKaplan1988}, and leave other observable signatures \cite{Arias2012,IrastorzaRedondo2018}. Common sources of such motion include misalignment from an initial displacement \cite{Preskill1983,Turner1986}, kinetic misalignment from a primordial velocity \cite{CoHallHarigaya}, and thermal trapping in a prescribed temperature-dependent potential \cite{BatellGhalsasi,DiLuzio2021}. A temperature-dependent rescaling of a single fixed breaking direction changes its magnitude but not its angular minimum; rotation requires differential reweighting of two or more non-aligned contributions. Here we show that differential thermal evolution of multiple non-aligned symmetry-breaking contributions across separated thresholds can instead rotate the effective pNGB vacuum direction and generate motion from rest at the instantaneous minimum.

Each explicit symmetry-breaking contribution defines a direction in the complex plane with a temperature-dependent magnitude \cite{DolanJackiw1974,Quiros1999,ArnoldEspinosa1993}. When two such contributions are non-collinear and are suppressed at different temperatures, one becomes subdominant before the other and the direction of their vector sum changes. We realize this effect in the Majoron sector of a Type-I seesaw mechanism, where two heavy Majorana neutrinos with separated thresholds supply the required structure \cite{DavidsonNardiNir,Buchmuller2005}. For a representative benchmark, the thermal breaking direction rotates by $|\Delta\theta_T|\simeq0.76$~rad and drives the Majoron to a peak velocity $|\dot\theta|/H\simeq0.62$, while the field tracks the moving minimum to sub-milliradian accuracy. Crucially, the time dependence of the minimum is not imposed externally; it emerges from differential thermal decoupling of states associated with the fixed, non-aligned breaking directions. This mechanism is not specific to the Majoron and can operate in other pNGB systems that have multiple non-aligned sources and distinct thermal evolution. In the Majoron realization, it provides a microscopic finite-temperature origin for the time-dependent background relevant to spontaneous leptogenesis.

\paragraph{Thermal torque.}
Geometrically, each explicit symmetry-breaking contribution is a vector in the complex plane, with a temperature-dependent magnitude and a fixed direction $\delta_i$. The angular potential is
\begin{equation}
V_{\rm ang}(\theta,T)
=
{\rm Re}
\left[
e^{i\theta}{\cal S}(T)
\right],
\qquad
{\cal S}(T)
=
\sum_i Q_i(T)e^{-i\delta_i},
\label{eq:S}
\end{equation}
where $Q_i(T)>0$ is the magnitude of each contribution and $\delta_i$ is its phase \cite{DolanJackiw1974,Quiros1999}. The instantaneous minimum sits at
\begin{equation}
\theta_{\min}
=
\pi-\arg{\cal S}.
\end{equation}
Entropy conservation in a radiation-dominated Universe gives
$\dot T/T=-\xi_T H$ with
$\xi_T=(1+\frac13\,d\ln g_{*s}/d\ln T)^{-1}$ so that
\begin{equation}
\frac{\dot\theta_{\min}}{H}
=
\xi_T\,
{\rm Im}
\left(
\frac{d\ln{\cal S}}{d\ln T}
\right).
\label{eq:master}
\end{equation}
For two contributions with relative phase
$\delta\equiv\delta_2-\delta_1$,
\begin{equation}
\frac{\dot\theta_{\min}}{H}
=
\xi_T
\frac{Q_1Q_2\sin\delta}{|{\cal S}|^2}
\left(
\frac{d\ln Q_1}{d\ln T}
-
\frac{d\ln Q_2}{d\ln T}
\right).
\label{eq:two}
\end{equation}
The minimum moves only if the breaking directions are non-collinear and their relative thermal weight evolves: $\sin\delta\neq0$ and $d\ln(Q_1/Q_2)/d\ln T\neq0$. We call this effect a thermal torque and define
\begin{equation}
\mathcal{T}_T
\equiv
{\rm Im}
\left(
\frac{d\ln{\cal S}}{d\ln T}
\right),
\qquad
\frac{\dot\theta_{\min}}{H}
=
\xi_T\,\mathcal{T}_T .
\end{equation}
For the benchmark below, $g_*=g_{*s}=106.75$ is constant, so
$\xi_T=1$.

Whether the pNGB follows the moving minimum is a separate question.
Adiabatic tracking requires $m_\theta\gg H$
\cite{DineRandallThomas1996,Kawasaki:2017xwt,Nakagawa2021}; in the
benchmark below, $m_\theta/H\gtrsim10^2$ throughout the
threshold-generated pulse. Adiabaticity therefore governs the field's
response to a moving minimum, not the origin of the motion. This is
what distinguishes thermal torque from the standard sources of pNGB
motion. Misalignment starts with a displaced field
\cite{Preskill1983,Turner1986}. Kinetic misalignment requires a
primordial angular velocity \cite{CoHallHarigaya}. Thermal trapping
relies on a temperature-dependent deformation of the potential
\cite{BatellGhalsasi,DiLuzio2021}. Adiabatic suppression mechanisms
prescribe the evolving minimum directly through the competing
potential \cite{Kawasaki:2017xwt}. Here the field begins at
$\theta_i=\theta_{\min}(T_i)$ with $\dot\theta_i\simeq0$, and its
motion is generated by the threshold-induced rotation of the minimum
itself.

\paragraph{Majoron realization.}
The Majoron sector of a Type-I seesaw mechanism provides a concrete realization
of thermal torque. We take the complex scalar and its potential
\begin{equation}
\Phi=
\frac{r}{\sqrt2}e^{i\theta},
\qquad
V_0=
\frac{\lambda_\Phi}{4}
(r^2-f_a^2)^2 ,
\end{equation}
with interactions
\begin{equation}
{\cal L}\supset
-\frac12
\sum_i
(\mu_i e^{i\delta_i}+y_i\Phi)N_iN_i
-
F_{\alpha i}\bar L_\alpha\widetilde HN_i
+{\rm H.c.}
\label{eq:L}
\end{equation}
The $\mu_i e^{i\delta_i}$ terms softly and explicitly break the global
lepton number, while $y_i\Phi$ gives the heavy neutrinos their masses
$M_{N_i}\simeq y_i f_a/\sqrt2$ once $r\simeq f_a$
\cite{Chikashige1980,Chikashige1981,GelminiRoncadelli,deGiorgi2024}.
Together with the Dirac Yukawas, these interactions realize the
Type-I seesaw mechanism and provide the lepton-number-violating structure
relevant for leptogenesis
\cite{FukugitaYanagida,DavidsonNardiNir,Buchmuller2005,Pilaftsis1997}.

Choose a basis with $y_i>0$ and define
$m_{0i}=y_i r/\sqrt2$. The Majorana mass squared contains the angular
interference term
$2\mu_i m_{0i}\cos(\theta-\delta_i)$. Each heavy neutrino therefore
contributes a breaking vector with a fixed direction $\delta_i$ and a
temperature-dependent weight. One contribution fixes a direction but
cannot rotate it; rotation requires at least two non-aligned
contributions whose thermal weights evolve differently. The two heavy
neutrinos provide precisely this structure.

To the first order in the angular interference, the thermal free energy
gives
\begin{equation}
A_i(T,r)
=
\frac{\mu_i m_{0i}T^2}{12}
{\cal F}_F
\left(
\frac{\bar m_i^2}{T^2}
\right),
\qquad
\ST
=
\sum_i A_i e^{-i\delta_i},
\label{eq:A}
\end{equation}
with $\bar m_i^2=\mu_i^2+m_{0i}^2$. The threshold function
${\cal F}_F$ follows from the finite-temperature effective potential
of the heavy Majorana states
\cite{DolanJackiw1974,Quiros1999} and is given explicitly in the
Supplemental Material. For the benchmark, the expansion parameters
$\epsilon_i\equiv2\mu_i m_{0i}/\bar m_i^2$ are
$\epsilon_1\simeq3.57\times10^{-3}$ and
$\epsilon_2\simeq1.43\times10^{-4}$. Replacing the linearized thermal
term by the exact finite-temperature angular determinant shifts the
peak angular velocity by less than $10^{-4}$, consistent with the
expected $\mathcal{O}(\epsilon_i^2)$ correction.

\paragraph{Zero-temperature contribution.}
The physical trajectory also depends on the renormalized
zero-temperature angular potential. We therefore define the total
complex breaking vector
\begin{equation}
\Bcal(T,r)
=
\Bvac(r)
+
\ST(T,r),
\label{eq:physicalS}
\end{equation}
where $\Bvac$ is obtained from the renormalized Coleman--Weinberg
potential \cite{ColemanWeinberg1973,Jackiw1974} and $\ST$ is the
thermal source of Eq.~\eqref{eq:A}. Along the temperature-dependent
radial background, we define
\begin{equation}
\frac{d}{d\ln T}\bigg|_{\rm bg}
\equiv
\frac{\partial}{\partial\ln T}
+
\frac{dr_{\rm bg}}{d\ln T}
\frac{\partial}{\partial r},
\label{eq:totalderiv}
\end{equation}
so that
\begin{equation}
\frac{\dot\theta_{\min}}{H}
=
\xi_T
\,{\rm Im}
\left[
\frac{
(d\Bcal/d\ln T)_{\rm bg}
}{
\Bcal
}
\right].
\label{eq:physicalvelocity}
\end{equation}
Here $\theta_{\min}(T)=\pi-\arg\Bcal(T,r_{\rm bg}(T))$, with the
sign in Eq.~\eqref{eq:physicalvelocity} following from
$\dot T/T=-\xi_T H$.

Temperature-independent local counterterms can modify $\Bvac$ and
hence the quantitative trajectory of the total minimum, but they do
not generate the differential thermal reweighting contained in
$\ST(T,r)$. The threshold-induced rotation is therefore a property
of the finite-temperature contribution, while the trajectory of the
total minimum depends on the renormalized zero-temperature potential,
corresponding renormalization conditions, and radial
evolution.

\paragraph{Threshold-generated rotation.}
We take a two-neutrino benchmark in which the separated masses and a
non-zero relative phase realize the two conditions in
Eq.~\eqref{eq:two}
\begin{equation}
\begin{gathered}
f_a=2.0\times10^{13}\ {\rm GeV},
\qquad
\lambda_\Phi=0.10,
\\
(y_1,y_2)=(0.20,1.00),
\qquad
\delta\equiv\delta_2-\delta_1=\pi/2,
\\
(\mu_1,\mu_2)
=
(5.0\times10^9,1.0\times10^9)
\ {\rm GeV}.
\end{gathered}
\label{eq:bench}
\end{equation}
The Yukawa hierarchy gives $M_{N_2}/M_{N_1}\simeq5$, while
$\delta=\pi/2$ provides strong non-collinearity.

The radial background is the broken-branch minimum of the leading
thermal-mass potential
\begin{equation}
r_{\rm LO}(T)
=
f_a
\sqrt{
1-
\frac{\Pi_\Phi(T)}
{\lambda_\Phi f_a^2}
},
\qquad
\Pi_\Phi(T)=\kappa_\Phi T^2 ,
\label{eq:rLO}
\end{equation}
with
$\kappa_\Phi=\lambda_\Phi/3+(y_1^2+y_2^2)/24$. Along this trajectory,
the threshold markers defined by
$\bar m_i(T_i^\star)/T_i^\star=1$ are
\begin{equation}
T_1^\star\simeq 2.81\times10^{12}\ {\rm GeV},
\qquad
T_2^\star\simeq 1.20\times10^{13}\ {\rm GeV}.
\end{equation}
As the two thresholds are crossed, the direction of the thermal
source $\ST(T,r)$ rotates, and the corresponding moving minimum
reaches
\begin{equation}
\max\frac{|\dot\theta_{T,\min}|}{H}\simeq0.486
\end{equation}
at $T_{\rm peak}^{(T)}\simeq6.76\times10^{12}$ GeV.

\begin{figure*}[t]
\centering
\includegraphics[width=0.96\linewidth]{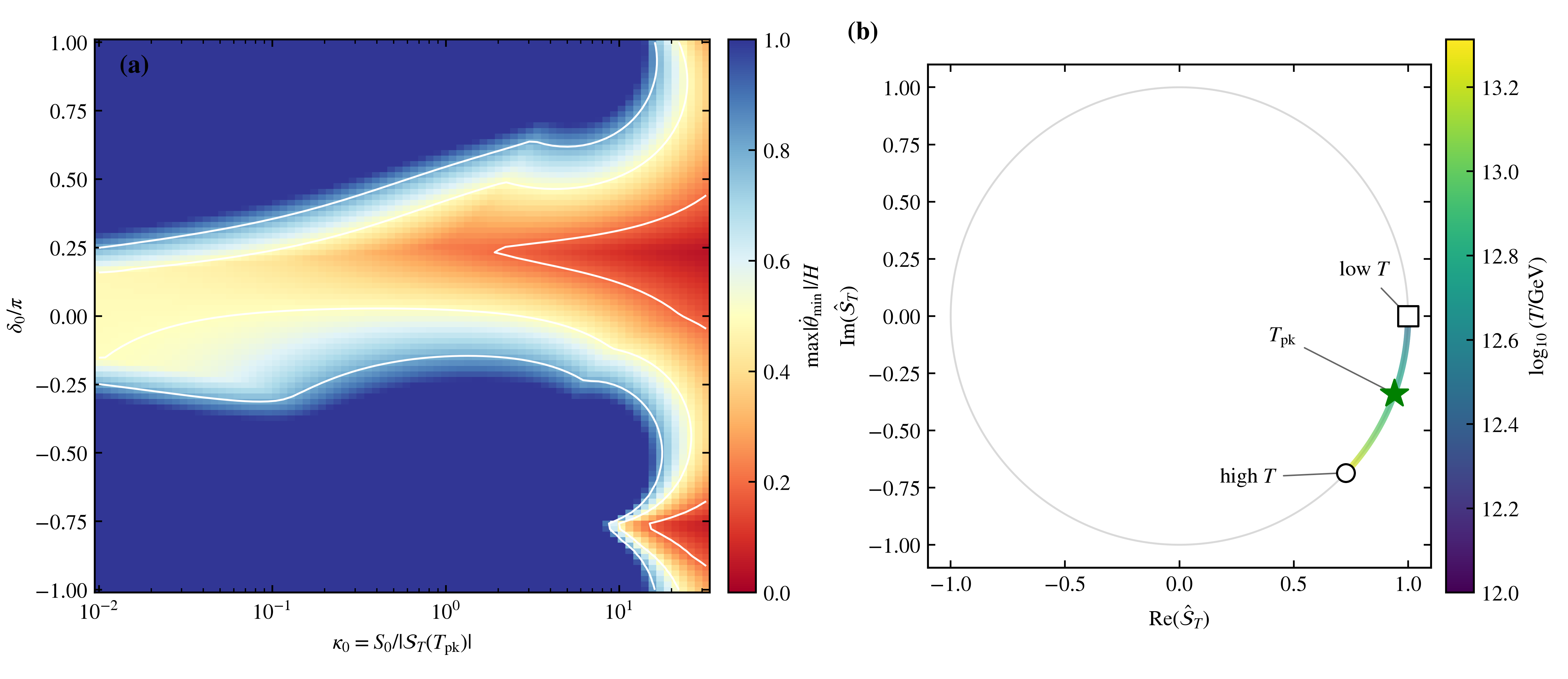}
\caption{\label{fig:thermal}
Thermal rotation and zero-temperature matching.
(a) Maximum moving-minimum velocity after adding a constant
zero-temperature breaking vector with magnitude
$\kappa_0|\ST(T_{\rm peak}^{(T)})|$ and orientation $\delta_0$.
The scan illustrates the dependence of the total trajectory on
zero-temperature matching without specifying a unique ultraviolet
completion.
(b) Direction of the thermal source,
$\hat{\mathcal S}_T=\ST/|\ST|$, shown on the unit circle during
cosmological cooling. The colour gradient denotes temperature. The
open circle and open square mark the high- and low-temperature
endpoints, respectively, and the green star marks
$T_{\rm peak}^{(T)}$. Their continuous angular separation is
$|\Delta\theta_T|\simeq0.76$~rad.}
\end{figure*}

To isolate this thermal rotation from zero-temperature matching, we
add a constant source,
\begin{equation}
{\cal S}_0^{\rm scan}=S_0e^{-i\delta_0},
\end{equation}
and define
$\kappa_0=S_0/|\ST(T_{\rm peak}^{(T)})|$.
Fig.~\ref{fig:thermal}(a) shows the maximum velocity of the total
instantaneous minimum as $\kappa_0$ and $\delta_0$ are varied, while
Fig.~\ref{fig:thermal}(b) isolates the thermal source by plotting
$\hat{\mathcal S}_T=\ST/|\ST|$ on the unit circle during cosmological
cooling.

The mechanism does not rely on finely balanced source strengths:
varying $0.3<{\cal R}_\mu=(\mu_1y_1)/(\mu_2y_2)<3$ preserves the
effect, with peak velocities spanning roughly a factor of five. A
broader scan over $1.15\le y_2/y_1\le5$ and
$0.04\pi\le\delta\le0.96\pi$ likewise shows the threshold-induced
rotation across a wide hierarchy--phase region whenever the angular
mode remains adiabatic.

\paragraph{From a rotating minimum to a moving field.}
A rotating instantaneous minimum generates physical Majoron motion
when the angular mode tracks the evolving potential adiabatically.
We solve the homogeneous field equation with the one-loop renormalized
angular potential \cite{ColemanWeinberg1973,Jackiw1974}, taking
$\mu_R=f_a$ and setting the finite angular counterterms to be zero at
this scale, which fixes the renormalization condition. The angular
equation is
\begin{equation}
\ddot\theta+
\left(
3H+2\frac{\dot r}{r}
\right)
\dot\theta
+
\frac{1}{r^2}
\frac{\partial V_{\rm eff}}{\partial\theta}
=0 .
\label{eq:eom}
\end{equation}

We consider the following initial field at the instantaneous matched minimum
\begin{equation}
\theta_i=\theta_{\min}(T_{\rm start}),
\qquad
\dot\theta_i=0,
\qquad
T_{\rm start}=0.90T_{c,0}^{\rm LO},
\label{eq:IC}
\end{equation}
where $T_{c,0}^{\rm LO}$ is the critical scale of the leading
thermal-mass minimum in the $\mu_i\to0$ limit. No initial displacement
or primordial velocity is imposed. As the minimum evolves, the
initially stationary field responds dynamically to the changing
potential. The motion reported below is therefore generated by the
evolving minimum, not by the initial condition.

For the benchmark, we have
\begin{equation}
\max\frac{|\dot\theta_{\min}|}{H}\simeq0.590
\end{equation}
at $T_{\rm peak}\simeq5.53\times10^{12}$ GeV, where
$r/f_a\simeq0.970$. At the peak, we have
$m_\theta/H\simeq3.3\times10^2$ and
$m_{\rho,\rm curv}/H\simeq2\times10^5$, while
$m_\theta/H\gtrsim10^2$ throughout the threshold-generated pulse.
The angular mode is therefore safely adiabatic
\cite{DineRandallThomas1996,Kawasaki:2017xwt,Nakagawa2021}, and the
Majoron tracks the moving minimum with
\begin{equation}
\max|\theta-\theta_{\min}|
\simeq
6.8\times10^{-4}\ {\rm rad},
\end{equation}
consistent with the analytic startup estimate
\begin{equation}
\frac{\bigl|
d\theta_{\min}/dN
\bigr|_{T_i}}
{(m_\theta/H)_i}
\simeq
6.7\times10^{-4}\ {\rm rad},
\end{equation}
where $N=\ln a$. The physical field velocity reaches
\begin{equation}
\max\frac{|\dot\theta|}{H}
\simeq0.62 .
\end{equation}
The small tracking transient reflects the initialization
$\dot\theta_i=0$ while the minimum is already beginning to move.

\begin{figure*}[t]
\centering
\includegraphics[width=0.99\linewidth]{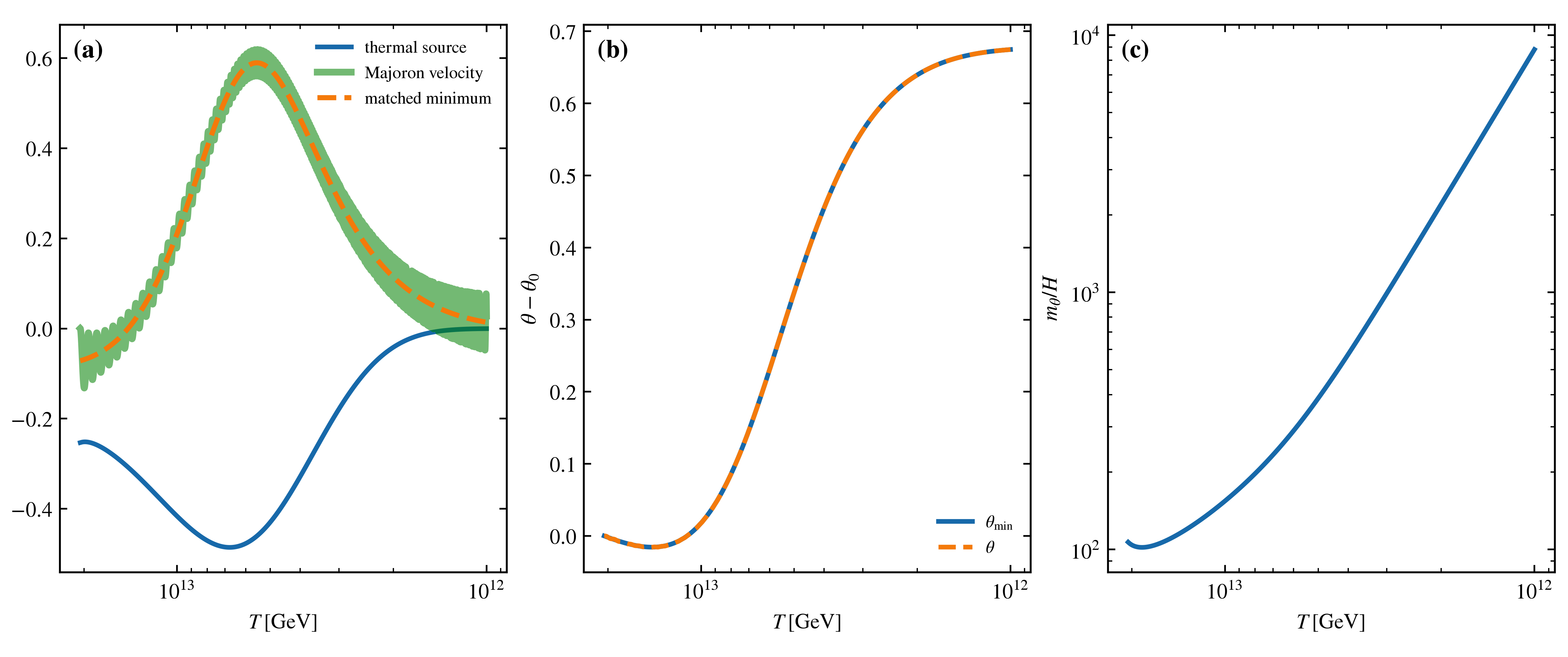}
\caption{\label{fig:dynamics}
Threshold-generated Majoron motion from the vanishing initial angular
velocity. (a) Thermal-only moving-minimum velocity, matched
moving-minimum velocity, and dynamical Majoron velocity from
Eq.~\eqref{eq:eom}. (b) Majoron field and instantaneous matched
minimum for $\dot\theta_i=0$. (c) Angular adiabatic hierarchy,
showing $m_\theta\gg H$ throughout the threshold-generated pulse.}
\end{figure*}

Figure~\ref{fig:dynamics} shows motion from rest: initialized at the
instantaneous matched minimum with $\dot\theta_i=0$, the Majoron reaches
$|\dot\theta|/H\simeq0.62$ as differential thermal reweighting moves the
minimum, with no imposed displacement or primordial velocity. Panel~(a)
compares the thermal-only, matched, and field velocities, panel~(b) shows
sub-milliradian tracking, and panel~(c) confirms $m_\theta\gg H$
throughout the pulse.

The resulting motion is the homogeneous field response to the
evolving effective potential, including the Hubble and radial damping
through Eq.~\eqref{eq:eom}. Additional dissipative effects and a full
non-equilibrium transport treatment are beyond the scope of this work.
The equilibrium-rate diagnostics and their dependence on the
heavy-neutrino Yukawa couplings are discussed in the Supplemental
Material. The generated background provides the microscopic input
for subsequent spontaneous-leptogenesis transport~\cite{ChunJungMajoron2024,ChunLeeSong2026}.

\paragraph{Connection to spontaneous leptogenesis.}
A time-dependent pNGB background can act as an effective chemical
potential that biases charge-violating reactions in the early
Universe and provides a central ingredient of spontaneous baryogenesis
and leptogenesis
\cite{CohenKaplan1987,CohenKaplan1988,IbeKaneta,ChaoPeng,
ChunJungMajoron2024,ChunLeeSong2026,Wada2024,King:2024idj}. In the
Majoron-seesaw realization, the differential thermal evolution of the
two heavy neutrinos supplies a microscopic finite-temperature origin
for such a background, generating a calculable homogeneous
$\dot\theta(T)$ without an imposed initial displacement or primordial
velocity. For the purely thermal benchmark, we have
\[
\frac{|\dot\theta_{T,\min}|}{T}
\simeq
4.61\times10^{-6}
\]
at
$T_{\rm peak}^{(T)}\simeq6.76\times10^{12}$~GeV, setting the
characteristic scale of the background that enters a subsequent
transport calculation.

This background alone does not determine the final baryon asymmetry.
Field redefinitions can redistribute the time-dependent phase among
mass terms, derivative interactions, and collision operators, so any
physical asymmetry requires a consistent and basis-independent transport
treatment \cite{Domcke2020,WashIn}. The thermal-torque mechanism
therefore supplies the dynamical pNGB background, while the resulting
lepton and baryon asymmetries depend on the subsequent transport
dynamics.

\paragraph{Conclusion.}
We have identified a finite-temperature mechanism that generates pNGB motion through differential thermal reweighting of multiple non-aligned symmetry-breaking sources. As heavy states cross separated thresholds, the thermal source direction rotates and drives a moving angular minimum. For the benchmark, the thermal source direction rotates by $|\Delta\theta_T|\simeq0.76$~rad, while the Majoron reaches a peak field velocity $|\dot\theta|/H\simeq0.62$ and tracks the moving minimum with sub-milliradian accuracy.

The Majoron-seesaw model provides a concrete realization in which two heavy neutrinos supply the required non-aligned breaking directions and separated thermal thresholds. Adiabaticity, $m_\theta\gg H$, determines whether the pNGB follows the moving minimum, but it is not the source of the motion. The differential thermal
reweighting itself is a finite-temperature effect, whereas the quantitative trajectory of the total minimum also depends on the zero-temperature matching and radial evolution. Thermal torque therefore provides a distinct microscopic origin for the pNGB motion alongside the misalignment, kinetic misalignment, and temperature-dependent trapping, and supplies a calculable time-dependent background for subsequent spontaneous-leptogenesis transport. The final lepton and baryon asymmetries require a basis-independent transport treatment, which we leave for a future work. More generally, the mechanism can operate whenever multiple non-aligned breaking sources undergo distinct thermal evolution; separated thresholds provide a simple microscopic realization.

\paragraph{Acknowledgments.}  
We would like to thank Xiaojun Bi and Yu Gao for valuable discussions, and our colleagues at the Institute of High Energy Physics and the Institute of Theoretical Physics for helpful comments. TL is supported in part by the National Key Research and Development Program of China Grant No. 2020YFC2201504, by the Projects No. 11875062, No. 11947302, No. 12047503, and No. 12275333 supported by the National Natural Science Foundation of China, by the Key Research Program of the Chinese Academy of Sciences, Grant No. XDPB15, by the Scientific Instrument Developing Project of the Chinese Academy of Sciences, Grant No. YJKYYQ20190049, by the International Partnership Program of Chinese Academy of Sciences for Grand Challenges, Grant No. 112311KYSB20210012, and by the Henan Province Outstanding Foreign Scientist Studio Project, No.GZS2025008. This work was supported by the High Performance Computing Platform of Henan Normal University.


\bibliographystyle{apsrev4-2}
\bibliography{references}


\section*{End Matter}

\paragraph{Numerical conventions.}
We solve the angular equation using $N=\ln a$ as the time variable,
with $\dot{\theta}/H=d\theta/dN$, from
$T_{\rm start}=0.90T_{c,0}^{\rm LO}$ to
$T_{\rm end}=10^{12}$ GeV. The benchmark uses the radial minimum
$r_{\rm LO}(T)$ of the leading-order thermal-mass potential and the
Hubble rate
$H(T)=1.66\sqrt{g_*}T^2/M_{\rm Pl}$, with
$M_{\rm Pl}=1.2209\times10^{19}$ GeV and
$g_*=g_{*s}=106.75$. The numerical field trajectory is used without
smoothing or post-processing. Contributions of thermally populated
beyond-SM degrees of freedom to the energy and entropy densities are
not included.

\paragraph{Validation scope.}
The threshold function ${\cal F}_F(x)$ is evaluated from converged
Bessel sums and cross-checked against direct quadrature at the
$10^{-8}$ level. The exact finite-temperature angular force is
computed from
$\partial V_T/\partial M^2=(T^2/24){\cal F}_F(M^2/T^2)$, avoiding
numerical differentiation of an interpolated thermal function. The
peak moving-minimum velocity is stable under refinement of the thermal
table at the $10^{-7}$ level, while the field solution is checked
against tighter ODE tolerances and doubled output sampling. These
checks establish numerical convergence within the physical
approximations used here; they do not independently validate the
radial, kinetic, or dissipative approximations beyond the tests
reported in the text.

\paragraph{Data availability.}
The numerical data and analysis code required to reproduce the figures
and benchmark results will be made publicly available in an online
repository upon publication.


\renewcommand{\theequation}{S\arabic{equation}}
\renewcommand{\thefigure}{S\arabic{figure}}
\renewcommand{\thetable}{S\arabic{table}}
\setcounter{equation}{0}
\setcounter{figure}{0}
\setcounter{table}{0}

\section{Finite-temperature determinant and threshold functions}
\label{sec:S-thermal}

We derive the finite-temperature contribution to the angular source
$\ST(T,r)$ used in the main text. For one Majorana
fermion, including its two physical fermionic degrees of freedom, the
one-loop thermal contribution is \cite{DolanJackiw1974}
\begin{equation}
V_T(r,\theta,T)
=
-\frac{T^4}{\pi^2}
\sum_i
J_F\!\left(
\frac{m_i^2(r,\theta)}{T^2}
\right),
\label{eq:S-VT}
\end{equation}
where
\begin{equation}
J_F(x)
=
\int_0^\infty \dd q\,
q^2
\ln\!\left(
1+e^{-\sqrt{q^2+x}}
\right).
\label{eq:S-JF}
\end{equation}
With this convention $J_F(x)>0$, decreases monotonically with $x$,
and satisfies $J_F(0)=7\pi^4/360$.

The field-dependent Majorana mass is
\begin{equation}
M_i(r,\theta)
=
\mu_i e^{i\delta_i}
+
\frac{y_i r}{\sqrt2}e^{i\theta}.
\label{eq:S-Mi}
\end{equation}
Working in the basis $y_i>0$, we define
\begin{equation}
m_{0i}
\equiv
\frac{y_i r}{\sqrt2},
\label{eq:S-m0}
\end{equation}
so that the physical mass entering the thermal determinant is
\begin{equation}
m_i^2(r,\theta)
=
\mu_i^2+m_{0i}^2
+
2\mu_i m_{0i}
\cos(\theta-\delta_i).
\label{eq:S-mi2}
\end{equation}

Introducing the angle-independent mass
\begin{equation}
\bar m_i^2
=
\mu_i^2+m_{0i}^2,
\end{equation}
the angular dependence can be expressed through the expansion
parameter
\begin{equation}
\epsilon_i
=
\frac{2\mu_i m_{0i}}{\bar m_i^2}.
\label{eq:S-eps}
\end{equation}
For the benchmark parameters and $r\simeq f_a$, the expansion
parameters are
\begin{equation}
\epsilon_1
\simeq
3.57\times10^{-3},
\qquad
\epsilon_2
\simeq
1.43\times10^{-4}.
\label{eq:S-eps-val}
\end{equation}
The angular interference is therefore perturbative, and the
first-order expansion used in Eq.~\eqref{eq:S-Ai} is parametrically
controlled; its accuracy relative to the full finite-temperature
determinant is verified numerically below.

\section{Expansion in the angular interference}
\label{sec:S-expansion}

Expanding the one-loop thermal potential in Eq.~\eqref{eq:S-VT}
around the angle-independent mass $\bar m_i^2$ gives
\begin{equation}
V_T(r,\theta,T)
=
V_T^{(0)}(r,T)
+
{\rm Re}
\left[
e^{i\theta}\ST(T,r)
\right]
+
{\cal O}(\epsilon_i^2),
\label{eq:S-VT-exp}
\end{equation}
where the temperature-dependent complex source is
\begin{equation}
\ST(T,r)
=
\sum_i
A_i(T,r)e^{-i\delta_i}.
\label{eq:S-ST}
\end{equation}
The individual thermal amplitudes are
\begin{equation}
A_i(T,r)
=
\frac{\mu_i m_{0i}T^2}{12}
{\cal F}_F
\left(
\frac{\bar m_i^2}{T^2}
\right).
\label{eq:S-Ai}
\end{equation}

The threshold function is obtained from the derivative of the
fermionic thermal function,
\begin{equation}
{\cal F}_F(x)
=
-\frac{24}{\pi^2}
\frac{\partial J_F(x)}{\partial x}.
\label{eq:S-FF}
\end{equation}
Because $\partial_xJ_F<0$, the threshold function satisfies
\begin{equation}
{\cal F}_F(x)>0,
\qquad
{\cal F}_F(0)=1.
\end{equation}
For $x\gg1$, it decreases exponentially as the corresponding heavy
state becomes thermally suppressed.

For numerical evaluation, we use the equivalent Bessel-$K$ expansion
\begin{equation}
{\cal F}_F(x)
=
\frac{12}{\pi^2}
\sum_{n=1}^{\infty}
(-1)^{n+1}
\frac{\sqrt{x}}{n}
K_1(n\sqrt{x}),
\label{eq:S-FF-series}
\end{equation}
which provides a numerically convenient representation of the
threshold function. In the numerical calculation, ${\cal F}_F$ is
tabulated on a logarithmic grid and evaluated using a cubic Hermite
spline whose nodal values and derivatives are obtained from the
converged Bessel sum.

The origin of the thermal rotation is Eq.~\eqref{eq:S-ST}. Different
heavy neutrinos acquire different thermal weights as their mass
thresholds are crossed. Their relative magnitudes therefore change
during cooling, rotating the direction of their vector sum in the
complex plane while the phases $\delta_i$ remain temperature
independent.

\section{Thermal rotation of the angular minimum}
\label{sec:S-rotation}

We now derive the motion of the instantaneous angular minimum from
the temperature-dependent complex source $\ST(T,r)$ of
Eq.~\eqref{eq:S-ST}. For the thermal-only case, the total breaking
vector coincides with the thermal source,
\begin{equation}
\Bcal=\ST .
\label{eq:S-Bthermal}
\end{equation}
Including the renormalized zero-temperature contribution gives the
total breaking vector
\begin{equation}
\Bcal(T,r)=\Bvac(r)+\ST(T,r),
\label{eq:S-Btotal}
\end{equation}
as used in the main text.

Writing
\begin{equation}
\Bcal(T,r)
=
|\Bcal(T,r)|e^{i\alpha(T,r)},
\end{equation}
the angular potential has the form
\begin{equation}
V_\theta
=
|\Bcal|
\cos(\theta+\alpha).
\end{equation}
Its instantaneous minimum is therefore
\begin{equation}
\theta_{\min}
=
\pi-\alpha
=
\pi-\arg\Bcal.
\label{eq:S-thetamin}
\end{equation}

Along a temperature-dependent radial background
$r=r_{\rm bg}(T)$, the total logarithmic derivative is
\begin{equation}
\frac{\dd}{\dd\ln T}\bigg|_{\rm bg}
=
\frac{\partial}{\partial\ln T}
+
\frac{\dd r_{\rm bg}}{\dd\ln T}
\frac{\partial}{\partial r}.
\label{eq:S-totalderiv}
\end{equation}
Entropy conservation gives
\begin{equation}
\frac{\dot T}{T}
=
-\xi_T H,
\qquad
\xi_T
=
\left(
1+
\frac13
\frac{\dd\ln g_{*s}}{\dd\ln T}
\right)^{-1}.
\label{eq:S-xi}
\end{equation}
Combining Eqs.~\eqref{eq:S-thetamin} and \eqref{eq:S-xi} yields the
angular velocity of the minimum,
\begin{equation}
\frac{\dot\theta_{\min}}{H}
=
\xi_T\,
{\rm Im}
\left[
\frac{
(\dd\Bcal/\dd\ln T)_{\rm bg}
}{
\Bcal
}
\right].
\label{eq:S-vmin}
\end{equation}

For two symmetry-breaking contributions,
\begin{equation}
\Bcal
=
Q_1e^{-i\delta_1}
+
Q_2e^{-i\delta_2},
\qquad
Q_i>0,
\end{equation}
with relative phase $\delta\equiv\delta_2-\delta_1$,
Eq.~\eqref{eq:S-vmin} reduces to
\begin{equation}
\frac{\dot\theta_{\min}}{H}
=
\xi_T
\frac{
Q_1Q_2\sin\delta
}{
|\Bcal|^2
}
\left[
\frac{\dd\ln Q_1}{\dd\ln T}
-
\frac{\dd\ln Q_2}{\dd\ln T}
\right].
\label{eq:S-two}
\end{equation}
For the benchmark we take $g_*=g_{*s}=106.75$ constant over the
displayed temperature range, so that $\xi_T=1$. The minimum moves only
if both conditions hold: $\sin\delta \neq 0$
and $\dd\ln(Q_1/Q_2)/\dd\ln T \neq 0$.

\section{Zero-temperature matching}
\label{sec:S-matching}

The thermal source $\ST(T,r)$ of Eq.~\eqref{eq:S-ST} does not by
itself constitute the complete angular potential; a zero-temperature
contribution from the same heavy Majorana neutrinos is also present.
The one-loop Coleman--Weinberg potential
\cite{ColemanWeinberg1973,Jackiw1974} in the $\overline{\rm MS}$
scheme is
\begin{equation}
V_{\rm CW}(r,\theta)
=
-\sum_i
\frac{
m_i^4(r,\theta)
}{
32\pi^2
}
\left[
\ln
\left(
\frac{m_i^2(r,\theta)}{\mu_R^2}
\right)
-
\frac32
\right].
\label{eq:S-CW}
\end{equation}
The coefficient $1/(32\pi^2)$ includes the Majorana fermionic
degree-of-freedom factor. We take the renormalization scale
\begin{equation}
\mu_R=f_a.
\label{eq:S-muR}
\end{equation}

Expanding Eq.~\eqref{eq:S-CW} to first order in the same angular
interference gives
\begin{equation}
V_{\rm CW}^{(1)}
=
\sum_i
C_i(r)
\cos(\theta-\delta_i),
\label{eq:S-CWlin}
\end{equation}
where
\begin{equation}
C_i(r)
=
-
\frac{
\mu_i m_{0i}\bar m_i^2
}{
8\pi^2
}
\left[
\ln
\left(
\frac{\bar m_i^2}{\mu_R^2}
\right)
-
1
\right].
\label{eq:S-Ci}
\end{equation}
This is the angular matching used in the baseline numerical
calculation.

The corresponding renormalized zero-temperature breaking vector is
\begin{equation}
\Bvac(r)
=
\sum_i
C_i(r)e^{-i\delta_i}.
\label{eq:S-Bvac}
\end{equation}
The physical angular minimum follows from the total breaking vector
$\Bcal=\Bvac+\ST$ introduced in Eq.~\eqref{eq:S-Btotal},
\begin{equation}
\theta_{\min}(T)
=
\pi
-
\arg
\Bcal
\left(
T,r_{\rm bg}(T)
\right).
\label{eq:S-thetamin-full}
\end{equation}
Temperature-independent local counterterms can modify $\Bvac$ and
hence the quantitative trajectory of the total minimum, but they do
not generate the differential thermal reweighting contained in
$\ST(T,r)$. The threshold-induced rotation is therefore a property
of the finite-temperature contribution, while the trajectory of the
total minimum depends on the renormalized zero-temperature potential,
the corresponding renormalization conditions, and the radial
evolution. The sensitivity to the renormalized zero-temperature
matching is illustrated by Fig.~\ref{fig:thermal}(a) of the main
text.

\section{Benchmark parameters}
\label{sec:S-benchmark}

The benchmark parameters are
\begin{equation}
\begin{gathered}
f_a = 2.0\times10^{13}\ {\rm GeV},
\qquad
\lambda_\Phi = 0.10,
\\
(y_1,y_2) = (0.20,1.00),
\qquad
(\mu_1,\mu_2) = (5.0\times10^9, 1.0\times10^9)\ {\rm GeV},
\\
\delta \equiv \delta_2 - \delta_1 = \frac{\pi}{2}.
\end{gathered}
\label{eq:S-bench}
\end{equation}

The baseline radial trajectory is the self-consistent minimum
$r_{\rm LO}(T)$ of the leading thermal-mass potential, defined in
Eq.~\eqref{eq:rLO} of the main text. Solving
\begin{equation}
\frac{\bar m_i(T_i^\star)}{T_i^\star}=1
\end{equation}
along this trajectory gives the characteristic threshold markers
\begin{equation}
T_1^\star
\simeq
2.81\times10^{12}\ {\rm GeV},
\qquad
T_2^\star
\simeq
1.20\times10^{13}\ {\rm GeV}.
\label{eq:S-Tstar}
\end{equation}
These are smooth crossover markers rather than physical phase
transition temperatures.

The principal baseline results are summarized in
Table~\ref{tab:S-bench}.

\begin{table}[t]
\centering
\caption{\label{tab:S-bench}
Benchmark parameters and baseline numerical results. Quantities with
a superscript $(T)$ correspond to the thermal source $\ST$ alone;
quantities without this superscript include the renormalized
zero-temperature contribution and use the baseline radial background
$r_{\rm LO}(T)$.}
\begin{ruledtabular}
\begin{tabular}{l l}
Quantity & Value \\
$f_a$ & $2.0\times10^{13}$ GeV \\
$\lambda_\Phi$ & $0.10$ \\
$(y_1,y_2)$ & $(0.20,1.00)$ \\
$(\mu_1,\mu_2)$ & $(5.0\times10^9,1.0\times10^9)$ GeV \\
$\delta$ & $\pi/2$ \\
$T_{c,0}^{\rm LO}$ & $2.2842\times10^{13}$ GeV \\
$T_1^\star$ & $2.81\times10^{12}$ GeV \\
$T_2^\star$ & $1.20\times10^{13}$ GeV \\
$T_{\rm peak}^{(T)}$ & $6.76\times10^{12}$ GeV \\
$\max|\dot\theta_{T,\min}|/H$ & $0.486$ \\
$T_{\rm peak}$ & $5.53\times10^{12}$ GeV \\
$\max|\dot\theta_{\min}|/H$ & $0.590$ \\
$r/f_a$ at peak & $0.970$ \\
$m_\theta/H$ at peak & $3.3\times10^2$ \\
tracking error & $6.8\times10^{-4}$ rad \\
\end{tabular}
\end{ruledtabular}
\end{table}

\section{Exact finite-temperature determinant validation}
\label{sec:S-exact}

We test whether the threshold-induced motion persists when the
linearized thermal expansion is replaced by the full finite-temperature
fermionic thermal potential.

At fixed radial background, the linearized thermal angular
contribution is replaced by
\begin{align}
\Delta V_T^{\rm exact}(r,\theta,T)
=
-\frac{T^4}{\pi^2}
\sum_i
\Bigg[
&
J_F
\left(
\frac{
\bar m_i^2
+
2\mu_i m_{0i}
\cos(\theta-\delta_i)
}{
T^2
}
\right)
\nonumber\\
&
-
J_F
\left(
\frac{\bar m_i^2}{T^2}
\right)
\Bigg].
\label{eq:S-dVT-exact}
\end{align}
The subtraction term is independent of $\theta$ and therefore does
not affect the location of the angular minimum.

The exact angular potential used in this comparison is
\begin{equation}
V_{\theta}^{\rm exact}
=
V_{\rm CW}^{(1)}
+
\Delta V_T^{\rm exact}.
\label{eq:S-Vtheta-exact}
\end{equation}
The linearized and exact calculations therefore use the same
zero-temperature matching prescription and the same radial background;
only the finite-temperature angular contribution is modified.

The exact angular force is evaluated from the identity
\begin{equation}
\frac{\partial V_T}{\partial M^2}
=
\frac{T^2}{24}\,
{\cal F}_F\!\left(\frac{M^2}{T^2}\right),
\label{eq:S-dVT-dM2}
\end{equation}
which follows from Eq.~\eqref{eq:S-FF}. This form avoids
differentiating an interpolated thermal function and provides the
numerically stable force used below.

The velocity of the instantaneous minimum is obtained from the
stationarity condition $F(\theta,T)\equiv\partial V/\partial\theta=0$
by implicit differentiation,
\begin{equation}
\frac{\dd\theta_{\min}}{\dd N}
=
-\frac{(\dd F/\dd N)_{\rm bg}}
{\partial F/\partial\theta},
\label{eq:S-implicit}
\end{equation}
which avoids numerical differentiation of a minimized angular array.
For the benchmark, the linearized and exact calculations give
moving-minimum peaks whose fractional difference is below $10^{-4}$.
The threshold-generated rotation therefore persists when the
linearized thermal expansion is replaced by the full thermal
potential. The residual is consistent with the expected
$\mathcal{O}(\epsilon_i^2)$ corrections for $\epsilon_i \sim 10^{-3}$.

\begin{figure}[t]
\centering
\includegraphics[width=0.9\linewidth]{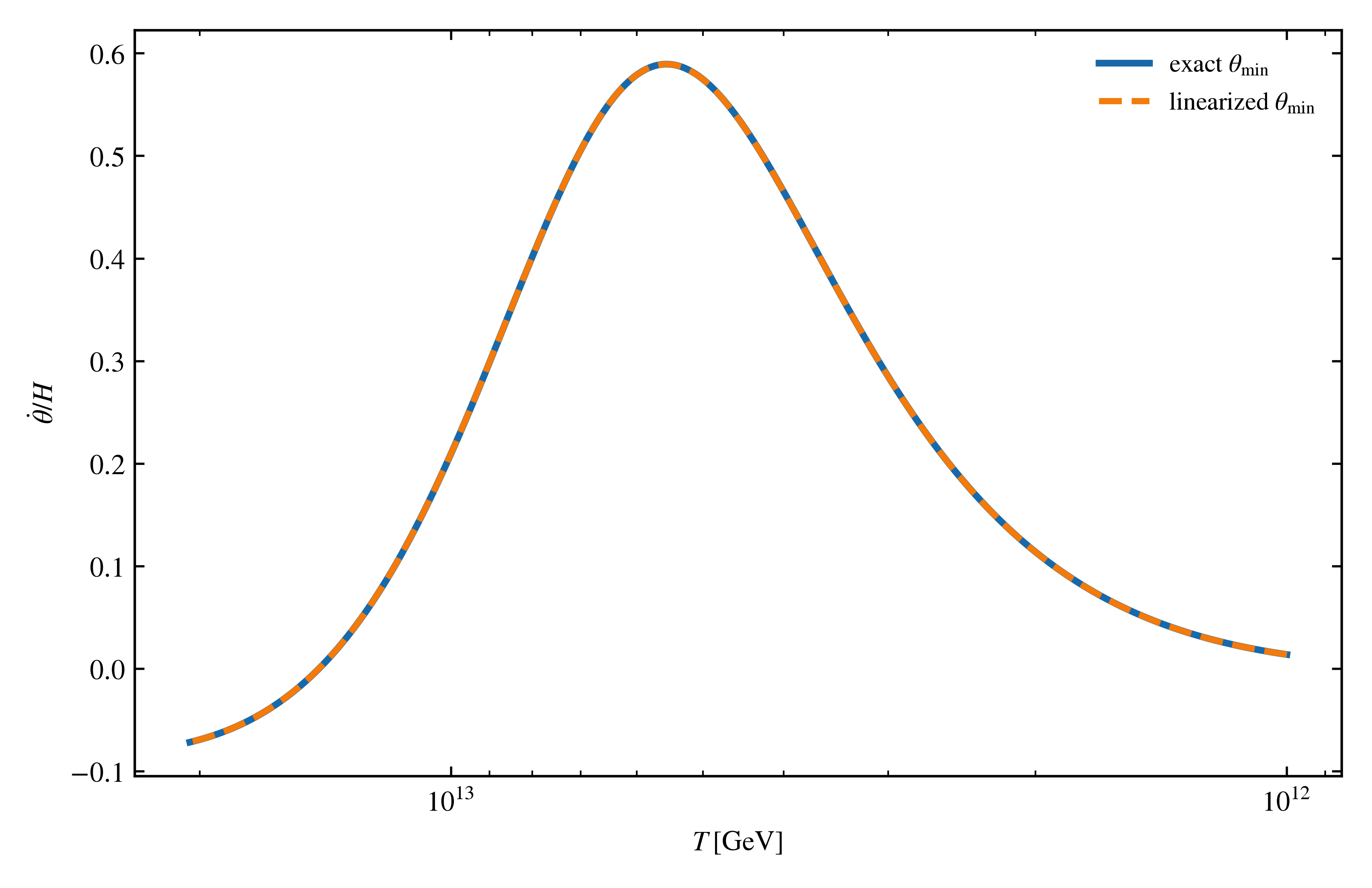}
\caption{\label{fig:S-exact}
Validation of the angular determinant expansion.
Moving-minimum velocity computed from the exact finite-temperature
angular force and from the linearized threshold expression. Both use
the same leading-order thermal-mass minimum $r_{\rm LO}(T)$ and the
same zero-temperature matching. The fractional difference between
the two curves remains below $10^{-4}$ over the threshold region,
consistent with the expected $\mathcal{O}(\epsilon_i^2)$ residual
of the angular-interference expansion.}
\end{figure}

\section{Radial background and sensitivity test}
\label{sec:S-radial}

\subsection{Baseline thermal-mass minimum}

The baseline calculation employs the minimum of the leading-order
thermal-mass potential,
\begin{equation}
V_{\rm LO}(r,T)
=
\frac{\lambda_\Phi}{4}(r^2-f_a^2)^2
+
\frac12 \Pi_\Phi(T) r^2,
\end{equation}
with
\begin{equation}
\Pi_\Phi(T)
=
\kappa_\Phi T^2,
\qquad
\kappa_\Phi
=
\frac{\lambda_\Phi}{3}
+
\frac{y_1^2+y_2^2}{24}.
\label{eq:S-kappa}
\end{equation}
Stationarity gives
\begin{equation}
r_{\rm LO}(T)
=
f_a\sqrt{1-\frac{\Pi_\Phi(T)}{\lambda_\Phi f_a^2}},
\label{eq:S-rLO-supp}
\end{equation}
which we use as the baseline radial trajectory. No ad hoc floor is
imposed. The leading-order expansion
$f_a\!\left(1-\Pi_\Phi/(2\lambda_\Phi f_a^2)\right)$ is the
small-$\Pi_\Phi$ limit of Eq.~\eqref{eq:S-rLO-supp} and is not used
as a baseline.

The critical scale in the $\mu_i\to0$ limit is
\begin{equation}
T_{c,0}^{\rm LO}
=
f_a
\sqrt{
\frac{
\lambda_\Phi
}{
\kappa_\Phi
}
}
=
2.2842\times10^{13}\ {\rm GeV}.
\label{eq:S-Tc}
\end{equation}
The baseline evolution begins at
\begin{equation}
T_{\rm start}
=
0.90T_{c,0}^{\rm LO}
=
2.0557\times10^{13}\ {\rm GeV}.
\end{equation}

\subsection{Threshold-aware refinement}

The fermionic contribution to the thermal scalar mass is suppressed
once the heavy states become non-relativistic. We therefore introduce
a threshold-aware radial trajectory $r_{\rm TA}(T)$, defined as the
minimum of the leading-order potential $V_{\rm LO}(r,T)$ with the
fermionic thermal mass evaluated including threshold suppression,
\begin{equation}
\Pi_\Phi^{\rm TA}(T)
=
\frac{\lambda_\Phi}{3}T^2
+
\sum_i
\frac{y_i^2 T^2}{24}\,
{\cal F}_F\!\left(
\frac{\bar m_i^2}{T^2}
\right),
\label{eq:S-PiTA}
\end{equation}
where $\bar m_i^2 = \mu_i^2 + m_{0i}^2$ is the angle-independent
fermion mass defined in Eq.~\eqref{eq:S-mi2}. At high temperature,
${\cal F}_F \to 1$ and Eq.~\eqref{eq:S-PiTA} reduces to
$\Pi_\Phi(T)$; as the heavy states decouple, the fermionic term is
Boltzmann suppressed and $r_{\rm TA}$ approaches $f_a$ from below
slightly faster than $r_{\rm LO}$. We use $r_{\rm TA}(T)$ as a
leading-order threshold-aware refinement of the radial background;
it is not presented as the complete one-loop finite-temperature
radial free energy.

\begin{figure}[t]
\centering
\includegraphics[width=0.9\linewidth]{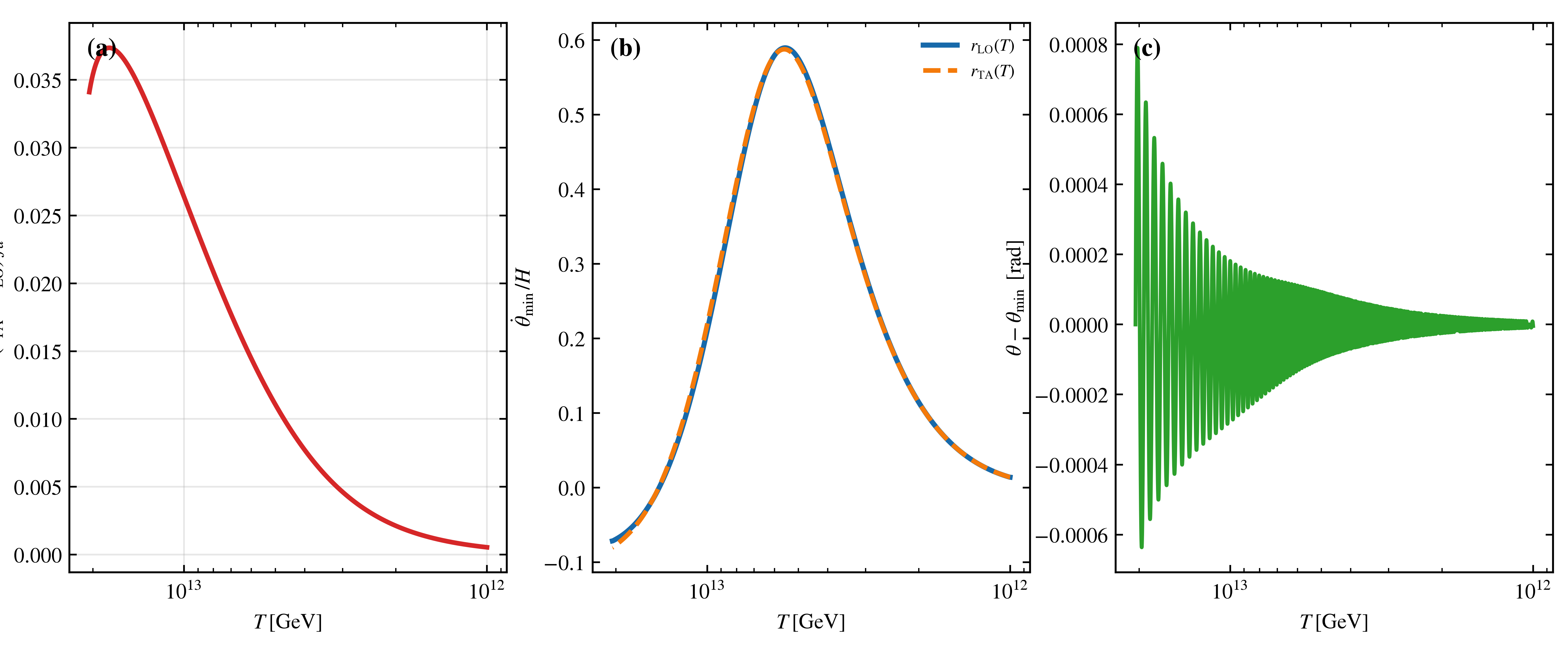}
\caption{\label{fig:S-radial}
Radial-background sensitivity test.
(a) Difference between the threshold-aware trajectory
$r_{\rm TA}(T)$ and the baseline self-consistent minimum
$r_{\rm LO}(T)$, normalized to $f_a$.
(b) Moving-minimum velocity on $r_{\rm LO}(T)$ (solid) and on the
threshold-aware refinement $r_{\rm TA}(T)$ (dashed), computed with
the same linearized angular potential. The two peak velocities differ
by $0.33\%$.
(c) Tracking error of the dynamical Majoron field relative to the
instantaneous minimum when the field equation is solved on the
threshold-aware trajectory. The oscillatory response remains
sub-milliradian, with a maximum deviation
$\simeq7.9\times10^{-4}$~rad.}
\end{figure}

We evaluate the moving-minimum velocity on both radial
trajectories using the same linearized angular potential, so that the
comparison isolates the effect of the radial background alone. The
two peak velocities agree to within $0.33\%$, as shown in
Fig.~\ref{fig:S-radial}. A static approximation $r=f_a$, which does
not respond to the thermal mass, suppresses the peak velocity by
$24\%$. The radial mode must therefore respond to the thermal background.
The difference between the two radial treatments tested here affects
the peak angular velocity only at the sub-percent level. The corresponding
field evolution remains adiabatic: the maximum
tracking error changes from $6.8\times10^{-4}$~rad on the baseline
trajectory to $7.9\times10^{-4}$~rad on the threshold-aware
trajectory, remaining safely below one milliradian.

\section{Numerical evolution of the Majoron field}
\label{sec:S-evolution}

The homogeneous angular equation governing the Majoron field is
\begin{equation}
\ddot\theta
+
\left(
3H
+
2\frac{\dot r}{r}
\right)
\dot\theta
+
\frac{1}{r^2}
\frac{
\partial V_{\rm eff}
}{
\partial\theta
}
=
0.
\label{eq:S-eom}
\end{equation}
Using $N=\ln a$ as the time variable, the second-order equation is
cast as a first-order system in the variables $(\theta,\omega)$,
where
\begin{equation}
\omega
\equiv
\frac{\dd\theta}{\dd N}
=
\frac{\dot\theta}{H}.
\end{equation}
The initial conditions are
\begin{equation}
\theta_i
=
\theta_{\min}(T_{\rm start}),
\qquad
\omega_i
=
0.
\label{eq:S-IC}
\end{equation}
This choice isolates the motion induced by the moving thermal
minimum; no initial angular displacement or primordial velocity is
imposed. The instantaneous minimum evolves with temperature, so the
field must respond dynamically as the Universe cools.

For constant $g_*$,
\begin{equation}
H(T)
=
1.66
\sqrt{g_*}
\frac{T^2}{M_{\rm Pl}},
\end{equation}
with
\begin{equation}
M_{\rm Pl}
=
1.2209\times10^{19}\ {\rm GeV},
\qquad
g_*=106.75.
\end{equation}
Equation~\eqref{eq:S-eom} is integrated numerically from
$T_{\rm start}$ down to $T_{\rm end}=10^{12}$ GeV using an adaptive
solver with relative and absolute tolerances of $10^{-10}$ and
$10^{-12}$, respectively. The field solution is checked against a
tighter tolerance run and against doubling the output sampling; the
peak velocity and maximum tracking error are stable at the $10^{-4}$
and $10^{-7}$ levels, respectively.

Using the leading-order thermal-mass minimum, the moving minimum
reaches
\begin{equation}
\max
\frac{
|\dot\theta_{\min}|
}{
H
}
\simeq
0.590
\end{equation}
at
\begin{equation}
T_{\rm peak}
\simeq
5.53\times10^{12}\ {\rm GeV},
\end{equation}
with $r/f_a \simeq 0.970$ and $m_\theta/H \simeq 3.3\times10^2$ at
this temperature.

The physical field velocity reaches
\begin{equation}
\max
\frac{
|\dot\theta|
}{
H
}
\simeq
0.62,
\label{eq:S-field-peak}
\end{equation}
close to the moving-minimum value. The small difference reflects the
forced oscillatory response of the field around the adiabatically
evolving minimum.

The maximum tracking deviation of the field from the instantaneous
minimum is
\begin{equation}
\max
|\theta-\theta_{\min}|
\simeq
6.8\times10^{-4}\ {\rm rad}.
\label{eq:S-track-base}
\end{equation}
This value matches the analytic startup-transient estimate
\[
|\theta_{\min}^{\prime}(T_i)|/(m_\theta/H)_i
\simeq
6.7\times10^{-4}\ {\rm rad},
\]
confirming that the observed deviation is the transient response to
initializing the field with $\dot\theta_i = 0$ while the minimum
begins to move, rather than a departure from adiabatic tracking.

\section{Adiabaticity}
\label{sec:S-adiabatic}

The physical angular curvature is defined by
\begin{equation}
m_\theta^2
=
\frac{1}{r^2}
\frac{
\partial^2V_{\rm eff}
}{
\partial\theta^2
}
\bigg|_{\theta=\theta_{\min}}.
\label{eq:S-mtheta}
\end{equation}
At the matched peak,
\begin{equation}
\frac{m_\theta}{H}
\simeq
3.3\times10^2.
\label{eq:S-mtheta-peak}
\end{equation}
The hierarchy $m_\theta \gg H$ is the standard mass-scale condition
supporting adiabatic following of a slowly evolving scalar minimum
in the early Universe
\cite{DineRandallThomas1996,Kawasaki:2017xwt,Nakagawa2021};
the rate of motion of the minimum and the field dynamics are
verified numerically in Sec.~\ref{sec:S-evolution}. In the benchmark
this hierarchy is satisfied throughout the threshold-generated pulse,
with $m_\theta/H \gtrsim 10^2$ over the full integration range, in
agreement with the small tracking error reported in
Sec.~\ref{sec:S-evolution}.

The radial-curvature estimate used in the benchmark satisfies
\begin{equation}
\frac{
m_{\rho,\rm curv}
}{
H
}
\sim
2\times10^5
\end{equation}
near the matched peak. Radial fluctuations are much faster than the
Hubble timescale there, supporting the treatment of $r(T)$ as a
prescribed background in the angular equation.

\section{Robustness under parameter variations}
\label{sec:S-robustness}

We test the robustness of the thermal torque away from the benchmark
configuration.

The benchmark value
\begin{equation}
\delta=\frac{\pi}{2}
\end{equation}
provides strong non-collinearity away from the aligned limits of
Eq.~\eqref{eq:S-two}; it is not, however, a required condition.

We scan the hierarchy and phase plane over the ranges
\begin{equation}
1.15
\le
\frac{y_2}{y_1}
\le
5,
\qquad
0.04\pi
\le
\delta
\le
0.96\pi.
\label{eq:S-scan-range}
\end{equation}
Regions with $m_\theta/H<10$, where adiabatic tracking is not
expected to be reliable, are masked in the scan.

\begin{figure}[t]
\centering
\includegraphics[width=0.9\linewidth]{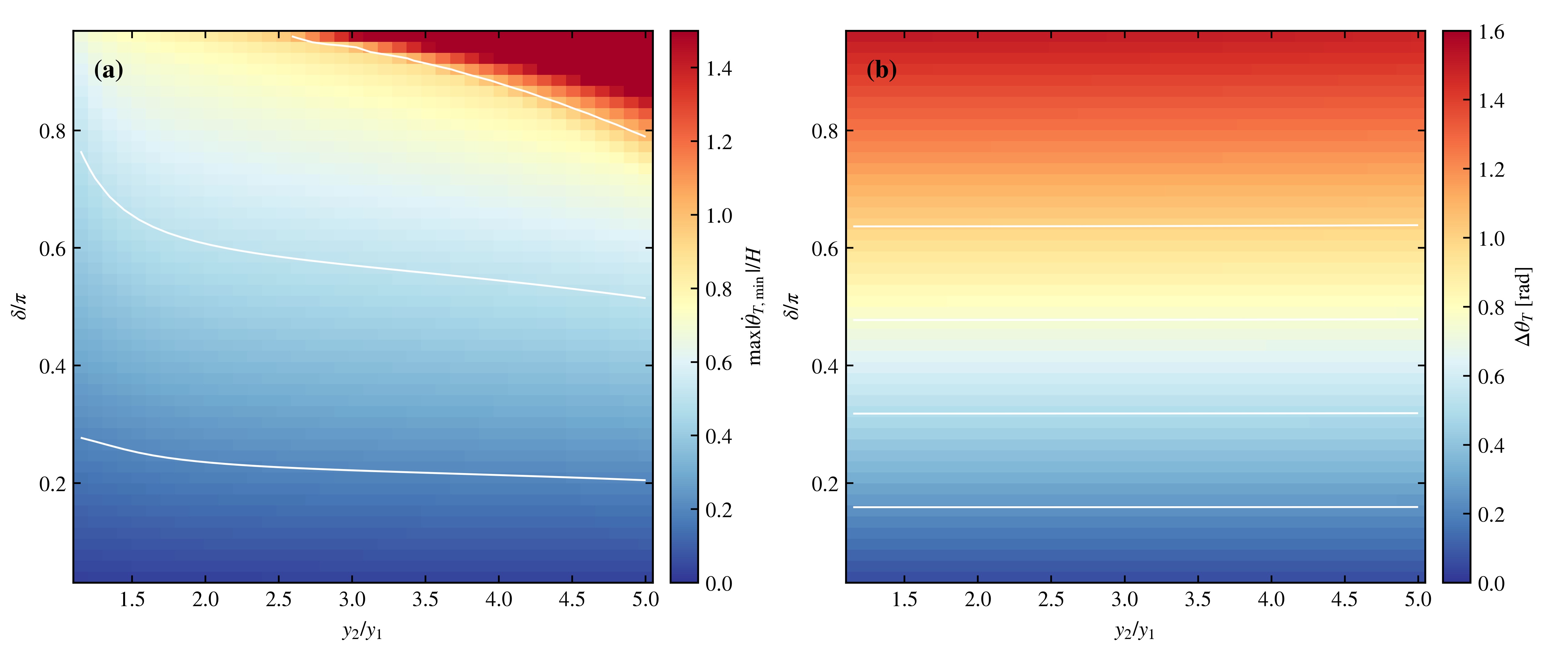}
\caption{\label{fig:S-thermal}
Thermal-torque robustness in the hierarchy--phase plane.
(a) Peak thermal moving-minimum velocity
$\max|\dot\theta_{T,\min}|/H$ as a function of the Yukawa hierarchy
$y_2/y_1$ and the relative phase $\delta/\pi$. Regions where the
angular hierarchy is too small for adiabatic tracking
($m_\theta/H<10$) are masked. The color scale is saturated at
$1.5$ for display; the largest unmasked value in the scan is
approximately $5.45$.
(b) Integrated thermal rotation $\Delta\theta_T$ over the same
parameter space. The benchmark of the main text corresponds to
$y_2/y_1=5$ and $\delta=\pi/2$.}
\end{figure}

We also vary the relative explicit-breaking strength
\begin{equation}
{\cal R}_\mu
=
\frac{
\mu_1y_1
}{
\mu_2y_2
}
\end{equation}
over
\begin{equation}
0.3
<
{\cal R}_\mu
<
3,
\end{equation}
which brackets the benchmark value ${\cal R}_\mu=1$ by a factor of
three in each direction. Across this interval the peak thermal
moving-minimum velocity spans
\begin{equation}
\max\frac{|\dot\theta_{T,\min}|}{H}
\simeq
0.21
\ \text{to}\
1.02,
\end{equation}
a variation of roughly a factor of five. The mechanism therefore
persists over an order-unity range of relative explicit-breaking
strengths, although its quantitative magnitude depends on the
relative strength of the two sources.

\section{Equilibrium rates and Yukawa benchmark}
\label{sec:S-kinetic}

The equilibrium calculation of the thermal source assumes thermal
heavy-neutrino distributions. To characterize the rate at which the
heavy neutrinos track the thermal bath, we use the Dirac Yukawa
benchmark
\begin{equation}
(F^\dagger F)_{11} = 2.0\times10^{-2},
\qquad
(F^\dagger F)_{22} = 1.374\times10^{-2},
\end{equation}
and evaluate
\begin{equation}
\frac{\Gamma_{D,i}}{H}
=
\frac{(F^\dagger F)_{ii}\,M_i}{8\pi H}\,
\frac{K_1(M_i/T)}{K_2(M_i/T)},
\label{eq:S-rate}
\end{equation}
with $M_i(T) = \sqrt{\mu_i^2 + m_{0i}^2(r_{\rm LO}(T))}$ evaluated
along the leading-order radial background. Near the matched peak this
gives
\begin{equation}
\frac{\Gamma_{D,1}}{H}\simeq 10,
\qquad
\frac{\Gamma_{D,2}}{H}\simeq 10^2.
\end{equation}
The decay and inverse-decay rates therefore support the equilibrium
thermal potential used to compute the angular source. Additional
interactions involving the Majoron-sector scalar can modify the
detailed kinetic relaxation and are not included in this diagnostic.
The values $(F^\dagger F)_{ii}$ are benchmark choices for this
equilibrium test. Compatibility with the observed light-neutrino
spectrum depends on the full Dirac Yukawa texture $F_{\alpha i}$,
which is not specified here. A kinetic treatment including the
scalar channel $\rho \to N_1 N_1$ is beyond the scope of this work.

\section{Numerical implementation}
\label{sec:S-numerics}

The baseline evolution runs from $T_{\rm start}$ down to
$T_{\rm end} = 10^{12}$ GeV with constant entropy degrees of freedom,
as described in Sec.~\ref{sec:S-evolution}.

The threshold function ${\cal F}_F$ is precomputed on a logarithmic
grid and evaluated by a cubic Hermite spline whose nodal values and
derivatives are obtained from converged Bessel sums, as described in
Sec.~\ref{sec:S-expansion}. This avoids repeated Bessel-sum evaluation
inside the ODE solver while preserving the derivative accuracy
required for the implicit differentiation of Eq.~\eqref{eq:S-implicit}.

The baseline angular equation of motion is integrated with the
\texttt{LSODA} method, using relative and absolute tolerances
\begin{equation}
{\tt rtol}=10^{-10},
\qquad
{\tt atol}=10^{-12}.
\end{equation}
The validation calculations quoted in Secs.~\ref{sec:S-exact} and
\ref{sec:S-radial} use independently tabulated exact thermal
functions. The field solution is checked against a tighter tolerance
run and against doubling the output sampling; the peak velocity and
maximum tracking error vary by less than $10^{-4}$ and $10^{-7}$,
respectively.

No smoothing is applied to the physical field trajectory. Peak
velocities are extracted directly from the numerical arrays rather
than from the rendered figures.

\section{Connection to spontaneous leptogenesis}
\label{sec:S-lepto}

The present calculation determines a homogeneous time-dependent
Majoron background but not the final lepton or baryon asymmetry.
Such a calculation requires a consistent transport description
including washout, collision terms, and the redistribution of the
time-dependent Majoron phase under field redefinitions
\cite{CohenKaplan1987,CohenKaplan1988,Domcke2020,WashIn,ChunLeeSong2026}.
For the thermal-only benchmark,
\begin{equation}
\max
\frac{
|\dot\theta_{T,\min}|
}{
H
}
\simeq
0.486
\end{equation}
at
\begin{equation}
T_{\rm peak}^{(T)}
\simeq
6.76\times10^{12}\ {\rm GeV}.
\end{equation}
Using
\begin{equation}
H(T)
=
1.66
\sqrt{106.75}
\frac{T^2}{1.2209\times10^{19}\ {\rm GeV}},
\end{equation}
this corresponds to
\begin{equation}
\frac{
|\dot\theta_{T,\min}|
}{
T
}
\simeq
4.61\times10^{-6}.
\label{eq:S-dot-over-T}
\end{equation}
The dimensionless ratio $|\dot\theta_{T,\min}|/T$ sets the
characteristic scale of the moving Majoron background entering a
subsequent transport calculation. Whether this background yields a
surviving asymmetry depends on the full Boltzmann dynamics, including
momentum-dependent collision terms and a basis-independent treatment
of the time-dependent phase. The final baryon asymmetry is therefore
left for future work.

\end{document}